\documentclass[reprint,superscriptaddress,showpacs,aps,prl]{revtex4-1}
\usepackage{amsmath,amssymb}

\usepackage{amsmath,amsfonts,amssymb}
\usepackage{graphicx}
\usepackage{algorithm}
\usepackage{algorithmic}
\usepackage{epstopdf}
\usepackage{float,placeins}
\usepackage[hyperfigures,breaklinks,colorlinks,citecolor=blue,linkcolor=black]{hyperref}
\usepackage[caption=false]{subfig}
\usepackage{mathbbol}

\begin{document}
\title{A theory of non-equilibrium local search on random satisfaction problems}

\author{Erik Aurell} 
\affiliation{Department of Computational Science and Technology, AlbaNova University Center, SE-106 91 Stockholm, Sweden}
\email{eaurell@kth.se}

\author{Eduardo Dom\'{\i}nguez}
\affiliation{Group of Complex Systems and Statistical Physics. Department of Theoretical Physics, Physics Faculty, University of Havana, Cuba}
\author{David Machado} 
\affiliation{Group of Complex Systems and Statistical Physics. Department of Theoretical Physics, Physics Faculty, University of Havana, Cuba}
\author{Roberto Mulet}
\affiliation{Group of Complex Systems and Statistical Physics. Department of Theoretical Physics, Physics Faculty, University of Havana, Cuba}
\email{mulet@fisica.uh.cu}

\date{\today}

\begin{abstract}
\noindent
We study local search algorithms to solve instances of the random $k$-satisfiability problem,
equivalent to finding (if they exist) zero-energy ground states
of statistical models with disorder on random hypergraphs.
It is well known that the best such algorithms are akin to non-equilibrium processes in a high-dimensional space.
In particular, algorithms known as focused, and which do not obey detailed
balance, outperform simulated annealing and related
methods in the task of finding the solution to
a complex satisfiability problem, that is to find (exactly or approximately) the minimum in a complex
energy landscape.
A physical question of interest is if the dynamics 
of these processes can be well predicted by the well-developed theory of equilibrium Gibbs states.
While it has been known empirically for some time that this is not the case,
an alternative systematic theory that does so has been lacking.     
In this paper we introduce such a theory based on the recently developed technique of cavity master equations
and test it on the paradigmatic random $3$-satisfiability problem.
Our theory predicts the solution process very accurately away from the algorithm phase
boundary and also predicts the qualitative form of this boundary.
\end{abstract}
\maketitle

\section{Introduction}
Combinatorial optimization problems
are of great importance
in many industrial and engineering fields,
and are also central to computational complexity~\cite{Garey1979}.
They are equivalent to the physical problem of finding 
ground states in statistical mechanics models with disorder, an analogy which has 
generated a large literature~\cite{MezardMontanari09}.
Constraint satisfaction is the subset of such problems where the energy function is
non-negative, and the problem is to find a zero-energy ground state (if any exists).
Many problems in combinatorial optimization are known to be worst-case computationally intractable, given ``$P\neq NP$''.

The typical or average-case behavior is however qualitatively different.
It was found a quarter of a century ago that the empirical run-time 
on random instances of one of the most famous NP-complete problems,
Boolean 3-satisfiability problem (3-SAT), varies greatly~\cite{Mitchell1992,statmechexamples1}. 
Very under-constrained problems are in practice easy, for 
almost any solution procedure.
This does not say that the problem would not be worst-case hard;
it only says that hard instances are hard to find in problems that are overall under-constrained.
The most difficult region is for problems that are on the verge of being
unsatisfiable, which for $3$-satisfiability means a ratio of clauses to variables ($M/N$)
of about $4.27$. Very over-constrained problems are again easy for complete algorithms, but
this aspect will not be discussed further here.

In the run-up from under-constrained to critical $3$-SAT problems different algorithms can be characterized,
rigorously or empirically, where they fail to work. We here take ``work'' to mean
``find a solution in time scaling polynomially in system size'', but keep it unspecified 
whether this has to happen always or with high probability, and leave 
aside rigorous considerations for which we refer to~\cite{AchlioptasCojaOhglan2008,CojaOghlan2016},
and references cited therein.
According to this criterion the best algorithm for random K-SAT is ``survey propagation''~\cite{statmechexamples2}
which in its most recent version is able to find solutions extremely close 
to the SAT/UNSAT threshold~\cite{Marino2016}.
Survey propagation is however a quite complex algorithm 
tailored to random constraint satisfaction problems, and is not competitive on most real-world problems~\cite{KaSe07}.
It is therefore of interest to step back and consider other 
simpler and more general solution procedures, of which the 
first example is ``simulated annealing''~\cite{SIM_ANNEAL},
a work-horse of scientific computing.
The performance of simulated annealing at slow enough cooling rate can be analyzed 
by spin glass techniques~\cite{Krzakala07} and is known to 
fail at some distance from the SAT/UNSAT threshold.
This can be taken to reinforce the (equilibrium) statistical mechanics view of random satisfiability problems.

The best local search algorithms that have been invented for satisfiability are however not
processes in detailed balance, and hence fall outside the paradigm of equilibrium statistical physics.
They all rely in ``focusing'', meaning that only variables that participate in some unsatisfied clause
are considered for update. A focused algorithm hence obeys to the dictum ``if it works, don't fix it''.
It is obvious that focusing breaks 
detailed balance as it leaves the set of solutions (zero-energy states) invariant.
In other words, if the problem has a solution,
then the focused algorithm has an absorbing set.
For constraint satisfiability, the most well-known algorithm in this class is ``walksat''~\cite{SKC} which
is competitive on many real-world problems~\cite{KaSe07,Barthel}.
Moreover, with parameter tuning it works on random $3$-SAT up to clause density about $4.2$~\cite{Gordon2005}.
Several other local search procedures have been shown to work up to a similar threshold~\cite{Schoning1999,Gordon2005,seitz05,Ardelius2006,alava07,Kroc2010,Lemoy2015}. We will here consider Focused Metropolis Search (FMS)~\cite{seitz05,alava07}.
This algorithm can be described very simply as first making a focusing step
and then a standard Metropolis step, as in simulated annealing at one temperature.
For the best choice of temperature FMS has been empirically shown to work up to clause density about $4.23$~\cite{seitz05}.

However, the understanding of non-equilibrium local search has been hampered by the absence of theory. While it has been empirically clear that predictions of equilibrium-derived theories do not apply, it has been unclear what to use in their stead.  The goal of this paper is to provide such a theory.  Previous attempts rest on average rate equations \cite{Barthel, SemerjianandWeigt} that must be built case by case. Our theory gives quantitatively excellent results on the development of FMS away from the SAT/UNSAT boundary, and qualitatively correct predictions on how that boundary depends on clause density and algorithm parameters. The crucial ingredient of this theory is the newly developed cavity method for continuous-time processes~\cite{AurellDelFerraroDominguezMulet2017}.

\section{Cavity Master Equation applied to random 3-SAT}
Cavity Master Equation is a closure of the dynamic cavity equations.
Dynamic cavity starts from the joint probability distribution of all 
histories of a set of dynamic variables interacting in a locally tree-like (locally loop-free) graph.
It is then possible to write a self-consistent equation for the
probabilities of the history of single variables when the history of one
of their neighboring variables is held fixed; one says that the first variable
is in the cavity of the second variable. These dynamic cavity equations are
formally  Belief Propagation updates. 
As is, they are however of little practical value since the variable (the history of
one dynamic variable) is very high-dimensional.
For dynamics in discrete time 
with synchronous updates closure assumptions have been explored for some time~\cite{NeriBolle,KanoriaMontanari,del2015dynamic}.

The Cavity Master Equation is appropriate for dynamics of discrete variables in continuous time.
In satisfiability problems these variables naturally take values $1$ (true) or $-1$ (false), which we here call spins. 
The Cavity Master Equation takes as input the jump rates $r_i$ (for spin $i$) defining the dynamics, and is
for spins interacting in groups labeled by $a,b,c,\ldots$ (constraints, clauses)
formulated in terms of quantities $p_{a\to i} (\sigma_{a\setminus i} | \sigma_i)$ where $\sigma_{a\setminus i}$
are the current values in group $a$ except $i$~\cite{P-spins}.
These quantities should be considered closures imposed on the corresponding full cavity quantities
$\mu_{a \to i} (X_{a\setminus i} | X_i)$ where $X_{a\setminus i}$ is the whole history of of all the spins in group
$a$ except $i$ in the cavity of $i$,
and $X_i$ the cavity history.
In practice, to describe FMS on random K-SAT we then have to solve the following set of coupled differential equations:
\begin{eqnarray}
\nonumber
 &&\dot{p}(\sigma_{a \setminus i} \! \mid \! \sigma_i) \! = \! - \!\! \sum_{j\in a\setminus i} \! \sum_{\substack{\lbrace \sigma_{b\setminus j }\rbrace \\ b \in \partial j \setminus a}} \!\!\! r_{j}(+) \!\!\!\!
 \displaystyle \prod_{b \in \partial j \setminus a} \!\! p( \sigma_{b\setminus j } \! \mid \! \sigma_{j})\;p(\sigma_{a \setminus i} \! \mid \! \sigma_{i})\\
 &&+ \sum_{j\in a\setminus i} \! \sum_{\substack{\lbrace \sigma_{b\setminus j }\rbrace \\ b \in \partial j \setminus a}} 
 \!\! r_{j}(-) \!\!\! \displaystyle \prod_{b \in \partial j \setminus a} \!\! p( \sigma_{b\setminus j } \! \mid \! -\sigma_{j})  \;p(F_j[\sigma_{a \setminus i}] \! \mid  \!\sigma_{i})
\label{eq:CME}
\end{eqnarray}
$F_j$ in above is the standard flip operator acting on spin $j$ while the
combination of several terms of the type $p(\sigma_{a \setminus i} \! \mid \! \sigma_{i})$
is characteristic of the cavity master equation closure, and structurally
analogous to the earlier described case of (ferromagnetic) $p$-spin model~\cite{P-spins}.
The term $r_{j}(\pm)$ in \eqref{eq:CME} 
is on the other hand the jump rate of spin $j$ when it takes value $\pm 1$. This quantity depends 
on the instantaneous value of spin $j$ and 
on the instantaneous values of all the spins interacting with $j$, through all the clauses in which spin $j$ appears. To describe the dynamics of the FMS algorithms one takes
\begin{equation}
r_i = \dfrac{E_i(\sigma_i, \sigma_{\partial i})}{K E} \text{min} \left[ e^{-\beta \Delta E(\sigma_i, \sigma_{\partial i})} , 1 \right]
\label{eqn:trans_rates_K_SAT}
\end{equation}
were $E_i(\sigma_i, \sigma_{\partial i})$ is the number of unsatisfied constraints of which spin $i$ is a member.
Each of these constraints can be written 
\begin{equation}
E_a = \dfrac{1}{2^{K}} \prod_{i \in a} \left( 1 - l_{i}^{a}\sigma_{i} \right)
\label{eqn:local_energy}
\end{equation}
$K$-satisfiability is thus a mixture of $p$-spin problems, where $p$ ranges from 1 to $K$.
FMS is based on focusing and a Metropolis step.
In the focusing of FMS all unsatisfied clauses are picked uniformly at random, 
and thereafter one variable in each such clause is again selected uniformly at random. This is the same as picking
all variables partaking in unsatisfied clauses with probability proportional to $E_i(\sigma_i, \sigma_{\partial i})$,
which explains this factor in~\eqref{eqn:trans_rates_K_SAT}.
The term $\text{min} \left[ e^{-\beta \Delta E(\sigma_i, \sigma_{\partial i})} , 1 \right]$
is on the other hand the standard Metropolis factor.

To model the dynamics of FMS in overall algorithmic time (wall-clock time),
we have to further take into account that the number of unsatisfied clauses 
changes. When this becomes smaller the rate per unit time of a given unsatisfied clause
to be picked goes up. This is reflected by the denominator $KE$ in
\eqref{eqn:trans_rates_K_SAT}, where $k$
is the number of variables per clause ($3$ for 3-SAT) and $E$ is the total number
of unsatisfied clauses. This factor kicks in strongly when there are only a few unsatisfied
clauses left, and when the variables in these clauses are probed more often. It can be eliminated
by letting the FMS algorithm mark time inversely proportionally to $E$, and is hence a kind of
globally defined time reparametrization.
By a more efficient coding one can bring down the number of sums in~\eqref{eq:CME}
from $2^{(K-1)c}$ to $2^{c}$ ($c$ is the number of clauses per variable). This coding is described in 
Supplemental Material

\section{Results}
The problem is defined by the ratio between the number of clauses ($M$) and the number of literals ($N$) of some given instance of 3-SAT written as $\alpha = M / N$, and by $\eta = e^{-\beta}$ as the noise parameter that enter into the rates of equation (\ref{eqn:trans_rates_K_SAT}). In order to understand the behavior of FMS we need to study its dependence on these two parameters.

\begin{figure}[H]
\centering
\includegraphics[keepaspectratio=true,width=0.35\textwidth]{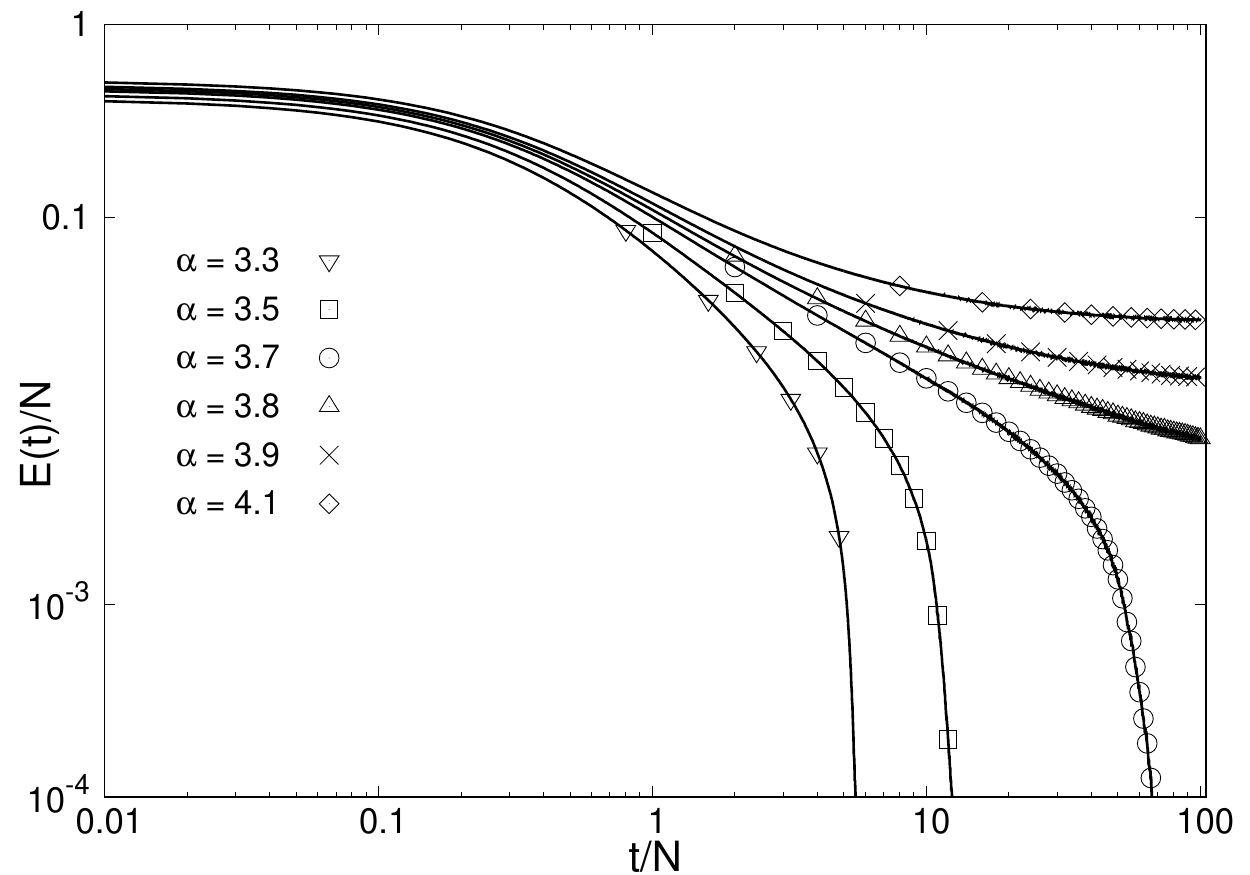}
\includegraphics[keepaspectratio=true,width=0.35\textwidth]{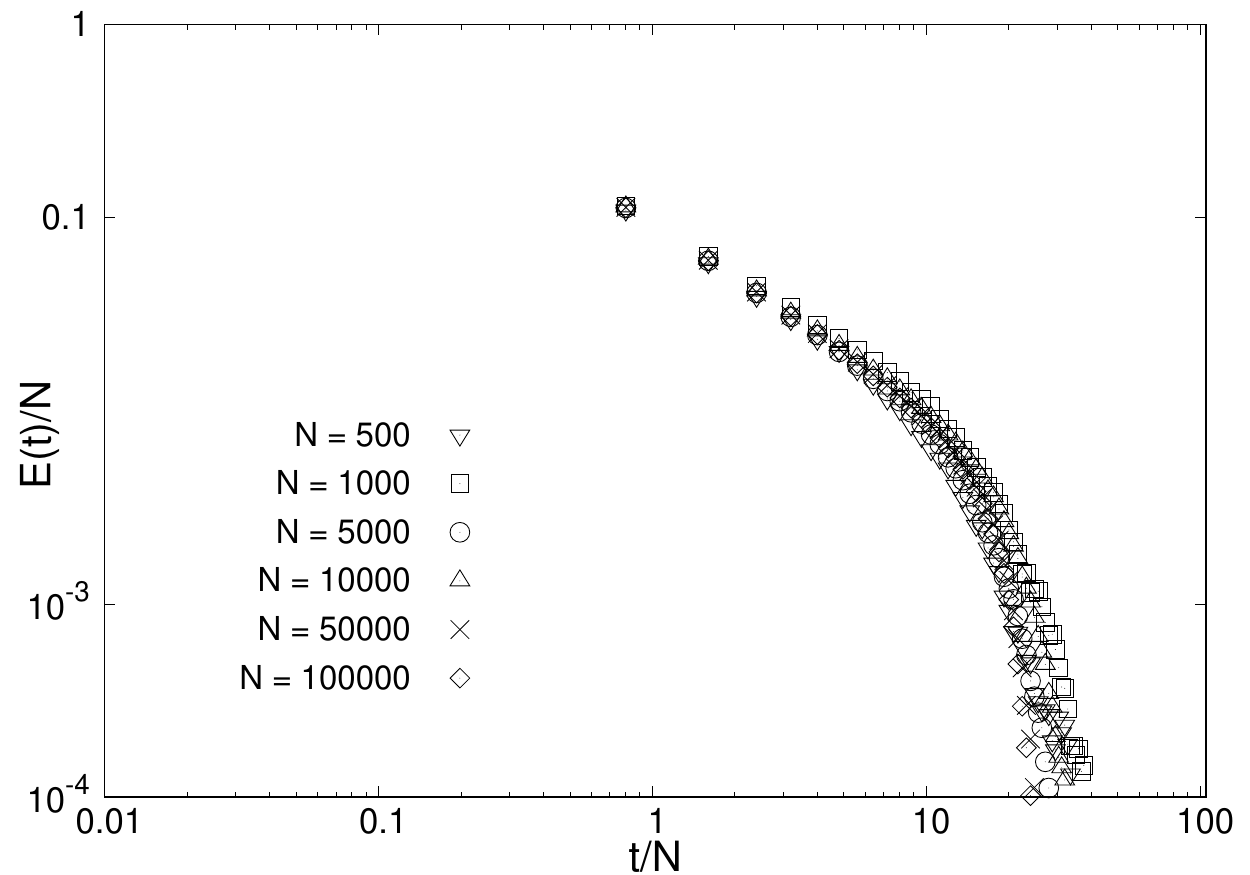}
\caption{FMS results on 3-SAT instances. Both panels show number of unsatisfied clauses (system energy) as function of time, in log-log scale. (Top) Dependency on $\alpha$ of FMS behavior. There is a transition between 
when FMS works (finds a solution in integration time considered), and when it does not. These calculations were done with $\eta = 0.45$ and system size $N=10^{5}$. Averages over 500 different histories were made for each $\alpha$. (Bottom) FMS's dependency on $N$ for $\eta=0.45$ and $\alpha=3.6 < \alpha_c$.}
\label{fig:FMS_dynamics}
\end{figure}

For a given noise $\eta$, FMS has been empirically shown to have a zone, for $\alpha$ lower than some $\alpha_c(\eta)$, where it solves 3-SAT instances in times linear with system size $N$. For $\alpha \geq \alpha_c$ solutions are found in times that grows exponentially with $N$, or solutions do not exist. This is shown in figure (\ref{fig:FMS_dynamics}). As can be seen in the top panel of this figure, for $\eta=0.45$, FMS is able to solve instances that have $\alpha \leq 3.7$, and seems to fail otherwise. In the bottom panel, size effects are represented. For $\alpha = 3.6$ FMS results seem to be almost independent of $N$.

\begin{figure}[H]
\centering
\includegraphics[keepaspectratio=true,width=0.35\textwidth]{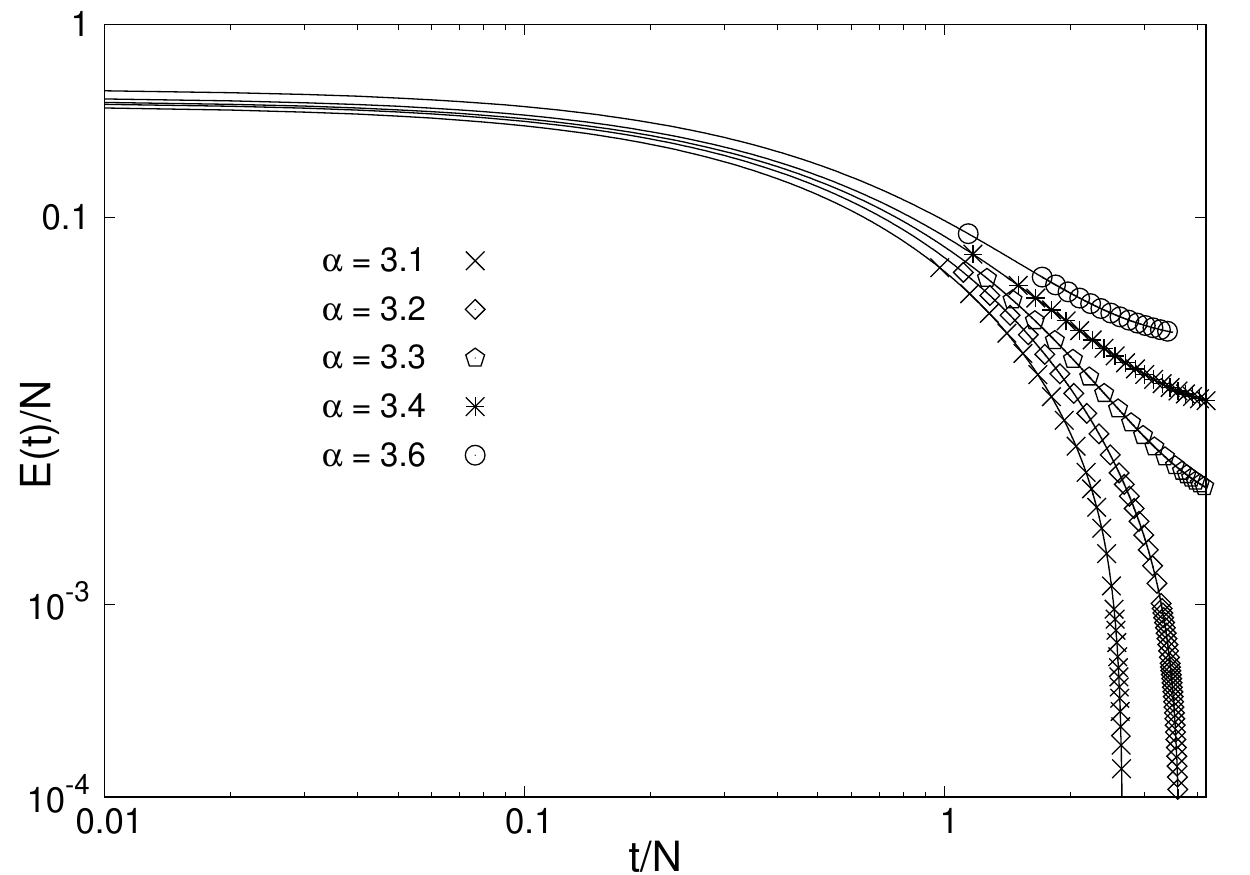}
\includegraphics[keepaspectratio=true,width=0.35\textwidth]{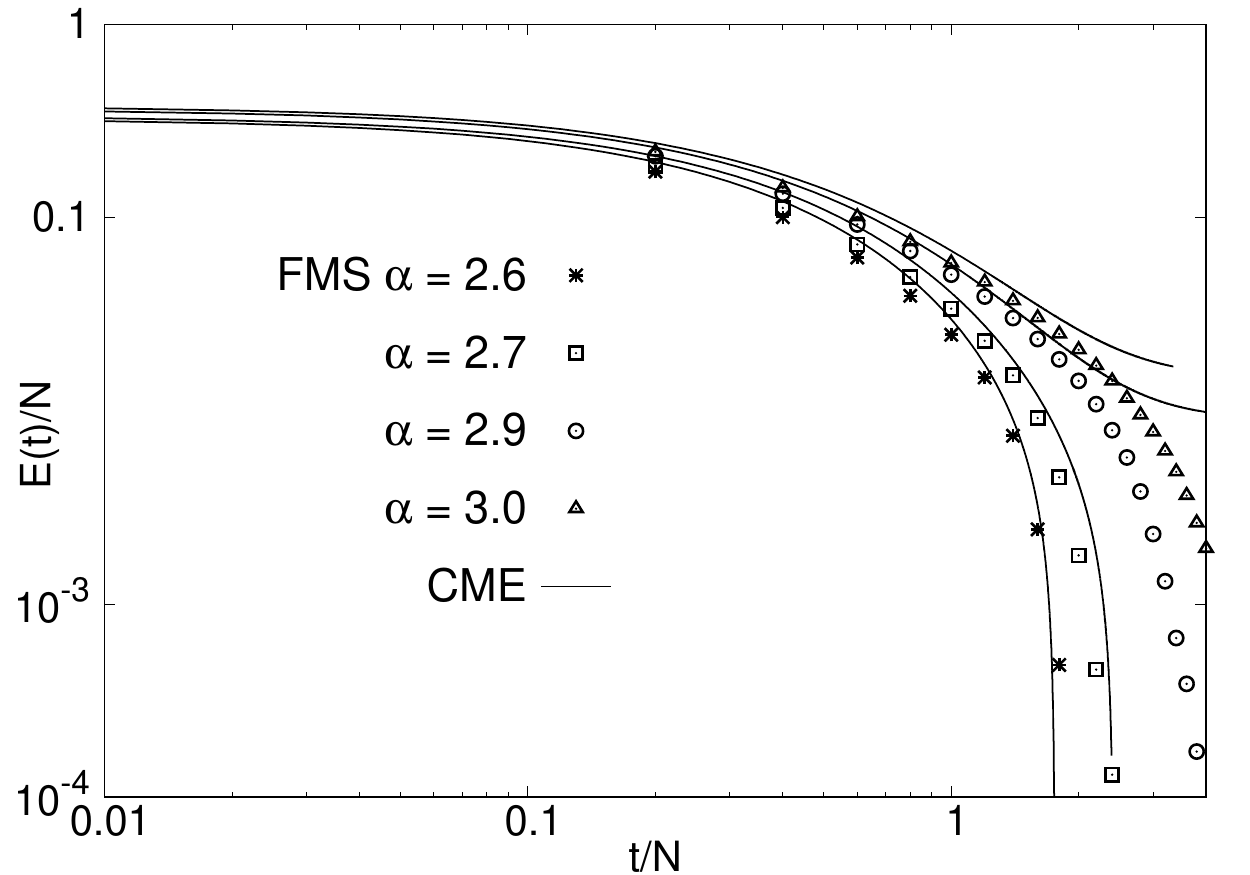}
\caption{CME results on 3-SAT instances. Both figures show the number of unsatisfied clauses (system energy)
as function of time, in log-log scale. (Top) Dependency on $\alpha$ of CME behavior. There is a transition between a phase in which
 solutions of CME reaches zero energy in finite time, and a phase where thhey do not. These calculations were done with $\eta = 0.45$ and system size $N=2000$. (Bottom) Comparison between CME (lines) and FMS (points) for $\eta=0.65$ and $N=5000$. In the region where frustration increases, i.e for high $\alpha$, CME will not reach zero energy even when FMS typically is able to find solutions. In this region, long range fluctuations 
in time and/or in the graph are  important.}
\label{fig:CME_dynamics}
\end{figure}

Then, by numerically integrating equations (\ref{eq:CME}) one can obtain the behavior for the same values of the parameter $\eta$. Results can be seen in figure (\ref{fig:CME_dynamics}, top). Although the transition $\alpha$ is not identical to figure (\ref{fig:FMS_dynamics}), the results of CME are qualitatively very similar. The differences are that CME, as is natural of the solution of a set of ordinary
differential equations, either converges to zero fairly rapidly, or does not converge to zero. The zone where FMS solves the problem by fluctuations is hence not well described by CME. The predicted threshold of CME ($\alpha_c$ for given $\eta$) is thus generally slightly smaller than the empirically determined threshold of FMS. 

On figure (\ref{fig:CME_dynamics}, bottom) a comparison is made between CME and FMS, for $\eta=0.65$, and several values of $\alpha$. Below the transition line of CME the agreement is very good. 

\begin{figure}[H]
\centering
\includegraphics[keepaspectratio=true,width=0.35\textwidth]{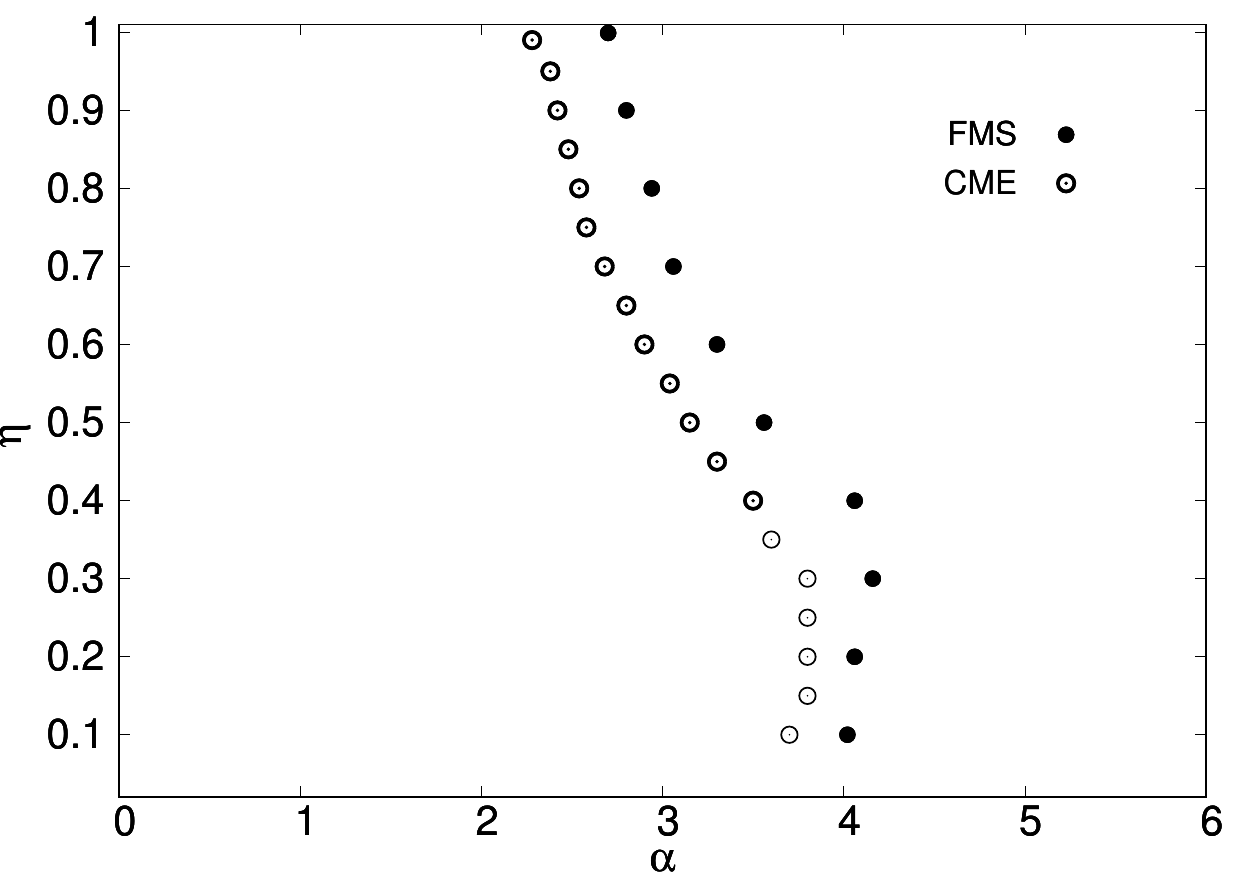}
\caption{Comparison between the phase diagrams of Focused Metropolis Search (FMS) and 
Cavity Master Equation (CME). The phase boundary of FMS was obtained by running $100$
instances of the problem at different $\alpha$ for a time $10^5\cdot N$ and determining at which value of $\alpha$
half of them finds a solution in the given time (filled-in circles). 
Convergence is slower in the lower ``descending'' branch of FMS.
The phase diagram of CME was found by integrating the CME for $24$ instances (thick-line circles) or at least $4$ instances (thin-line circles) of the problem at different $\alpha$. Results for each $\alpha$ were placed in log-log plots as the ones of figure (\ref{fig:CME_dynamics}) and it was determined for which value of $\alpha$ half of them did not converged to zero.}
\label{fig:CME_vs_FMS_phase_diag}
\end{figure}

As a summary, a comparison between the corresponding phase diagrams of FMS and CME is shown in figure (\ref{fig:CME_vs_FMS_phase_diag}). 
As one sees there is high qualitative similarity between them, essentially the transition line in CME is pushed to
a little smaller values of $\alpha$ but the two curves follow each other quite closely
as the parameter $\eta$ is varied.  

\section{Discussion}
The qualitative and quantitative description of the energy landscapes 
in combinatorial optimization problems is one of the most important results
of statistical physics of disordered systems, with many applications in many
areas of science~\cite{MezardParisiVirasoro,MezardMontanari09}.
The quantitative prediction of the exact threshold between 
a SAT and an UNSAT phase in random satisfiability problems 
by a one-step replica symmetry breaking (1RSB)
technique was a breakthrough~\cite{statmechexamples2}, which
has been extended to many other paradigmatic problems in computer science such
as \textit{e.g.} graph coloring~\cite{Mulet2002}
vertex covering~\cite{Zhou2009}, and the stochastic block model~\cite{Decelle2011}.

Yet, these advances a priori describe statics, and not dynamics.
A long line of empirical investigations surveyed in the introduction have shown 
that the phase diagram of non-equilibrium local search appears unrelated to 
bounds derived from the complexity
of (equilibrium) free energy landscapes.
A further and more recent discussion that non-equilibrium may be ``unreasonably effective''
was given in \cite{Baldassi2016} and similarly in \cite{Budzynski}.
For combinatorial optimization it may hence be possible to achieve  
what has sometimes been posited to be impossible, from equilibrium considerations.
A full realization (and exploitation) of these results has however been hampered by a lack of systematic
theory. This is what we have furnished here, by adapting recent advances in the description
of dynamics on locally tree-like graphs.

Our theory for how the local search proceeds in time
is very accurate away from the (algorithm-dependent) phase boundary.
The discrepancies found close to the phase boundary are very likely due
to the build-up of correlations in time which are not captured by the closure
approximation that leads to the Cavity Master Equation.
We note that in the simpler case of synchronously updated spin systems (parallel updates)
it was possible to improve on an analogous Markov approximation 
presented in~\cite{del2015dynamic} by using
the matrix product approximation of quantum many-body theory~\cite{barthel2018matrix}. We believe it is likely that efficient and more accurate
higher-order closure schemes can also be found for continuous-time dynamics.
For Focused Metropolis Search we find that our theory 
captures well the form of the phase boundary: for given $\eta$ (Metropolis parameter)
the predicted boundary is basically shifted to a somewhat smaller value of $\alpha$ (clause density).  

\section*{Acknowledgments}
We acknowledge support from the European Union Horizon 2020 research and innovation  programme  MSCA-RISE-2016  under  grant agreement  No. 734439 INFERNET and by an Erasmus+ International Credit Mobility to KTH (EU).

\bibliographystyle{apsrev4-1}
\bibliography{KSAT_joined}

\end{document}